\documentstyle[11pt,newpasp,twoside,psfig]{article}
\begin{document}

\title{University of St.~Andrews Open Cluster Survey for Hot Jupiters}
\author{R.~A. Street$^1$, Keith Horne$^2$, A. Collier Cameron$^2$, Y.~Tsapras$^3$,
D.~Bramich$^2$, A. Penny$^4$, A. Quirrenbach$^5$, N. Safizadeh$^5$, 
D. Mitchell$^5$, J.~Cooke$^5$}
\affil{$^1$APS Division, School of Physics, Queen's University of Belfast,
University Road, Belfast, BT7 1NN, Northern Ireland,\\
$^2$School of Physics and Astronomy, University of St. Andrews, North Haugh,
St. Andrews, Fife, KY16 9SS, Scotland, \\
$^3$School of Mathematical Sciences, Queen Mary University of London, 
Mile End Road, London, E1 4NS, England, \\
$^4$Rutherford Appleton Laboratories, Chilton, Didcot, Oxon, OX11 0QX,
England,\\
$^5$Center for Astrophysics and Space Sciences (CASS), University of California, 
San Diego, CA 92093, USA.  }

\begin{abstract} 

We are using the Isaac Newton Telescope Wide Field Camera to survey open
cluster fields for transiting hot Jupiter planets.  Clusters were selected on
the basis of visibility, richness of stars, age and metallicity.  Observations
of NGC 6819, 6940 and 7789 began in 1999 and continued in 2000.  We have
developed an effective matched-filter transit-detection algorithm which has
proved its ability to identify very low amplitude eclipse events in real data. 
Here we present our results for NGC 6819.  We have identified 7 candidates
showing transit-like events.  Colour information suggests that most of the
companion bodies are likely to be very-low-mass stars or brown dwarfs,
intrinsically interesting objects in their own right.  

\end{abstract}

\section{Introduction}

The technique of discovering planets from their photometric transits enables us
to probe fainter stars for planetary companions than are accessible to radial
velocity surveys.  This allows us to probe a wider variety of stellar
environments.  The technique also offers us the possibility of discovering
statistically significant numbers of hot Jupiter planets from the ground within
a relatively short timescale.  Such a sample could be used to establish the
relationships between planetary formation and key properties such as
metallicity, age, etc. Janes (1996) suggested that clusters would be ideal
targets for a transit search.  They provide large numbers of stars within a
relatively small area, the ages and metallicities of which can be determined. 
Radial velocity (RV) surveys have indicated that stellar metallicity in
particular seems to be correlated with the presence of planets.  Brown et al.
(2001) reported surveying the globular cluster 47 Tuc with the Hubble Space
Telescope, but interestingly found no transits.  In the light of the RV
results, this raises the question of whether and how this deficit is related to
the low metallicity of the cluster and/or star/radiation density.  To address
this question, we are surveying a number of open clusters with a range of
ages/metallicities.  Open clusters do not suffer from the dense crowding of
stars present in globulars and so can be studied with ground-based telescopes. 
They are also generally younger and higher in metallicity.  

\section{Observations and Data Reduction} 

\begin{figure*}
\centering
\begin{tabular}{cc}
\psfig{file=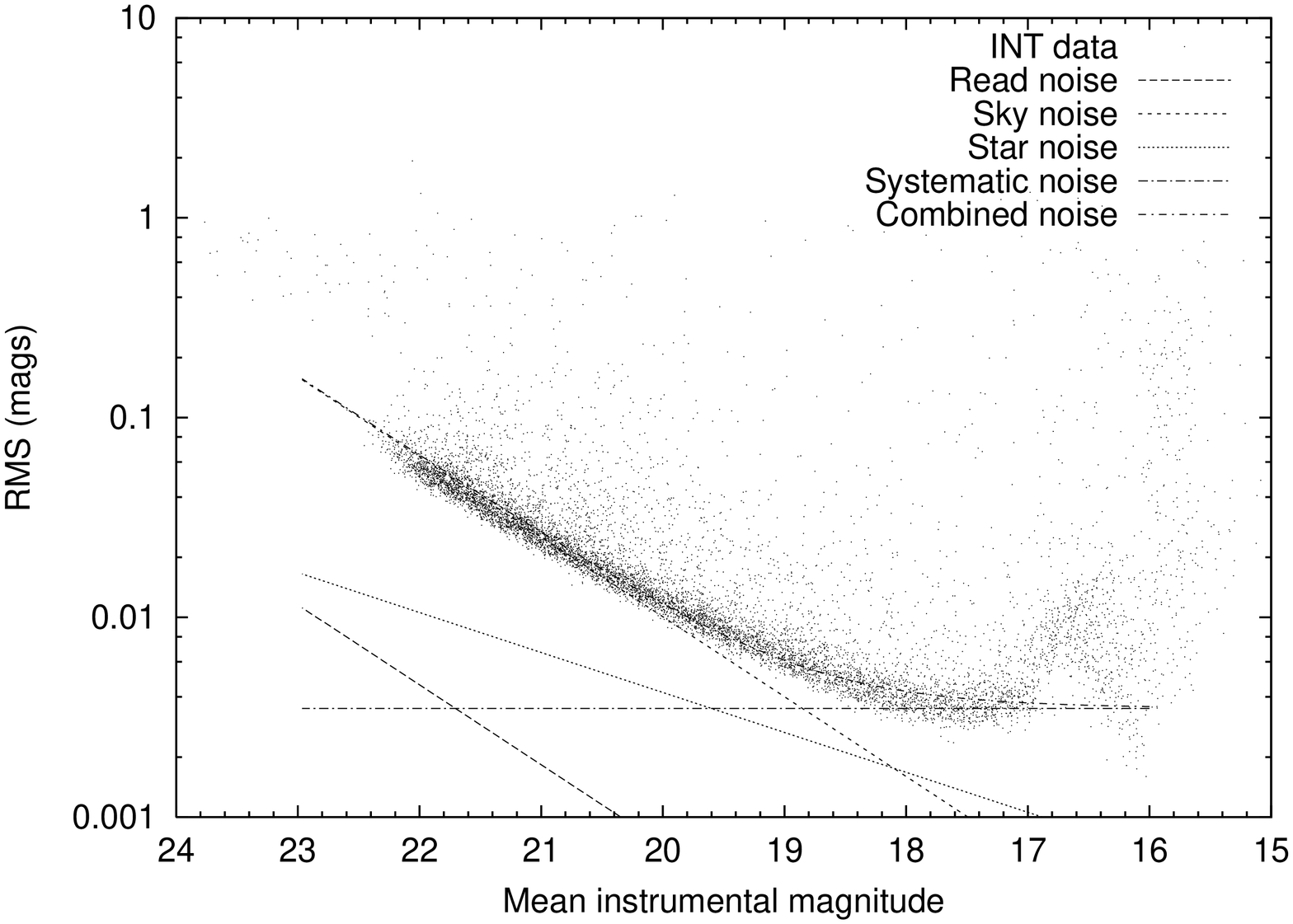,angle=270.0,width=6cm}
&
\psfig{file=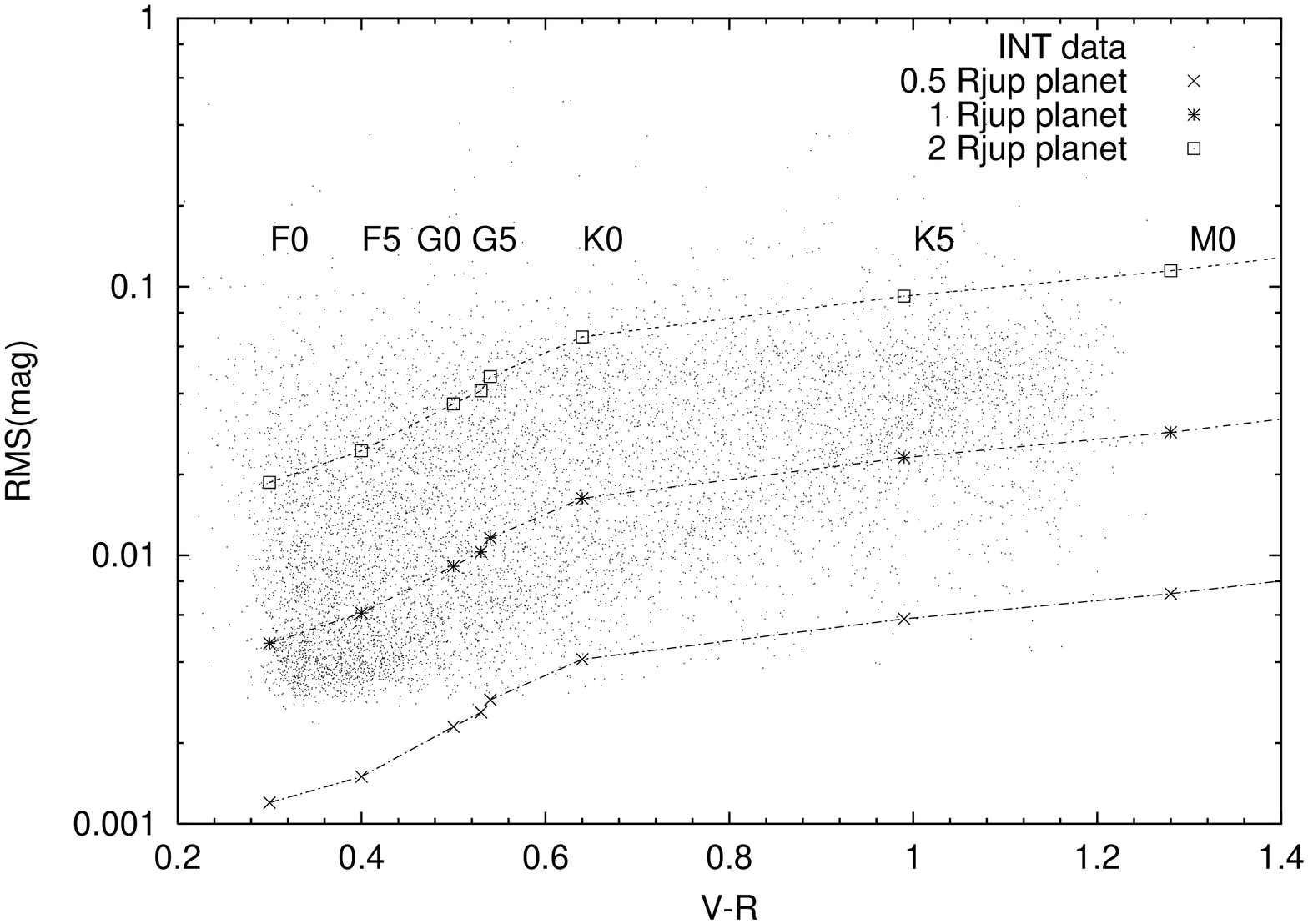,angle=270.0,width=6cm} \\
\end{tabular}
\caption{RMS scatter of each lightcurve plotted against instrumental mean
magnitude (left) and V-R (right) for CCD4.  Dots indicate predicted depth of
transit caused by 1 $R_{jup}$ radius planet orbiting a cluster-member star of 
given spectral type.  We can detect such planets in lightcurves of majority of
stars.  }
\protect\label{fig:RMSfigures}
\end{figure*}

We were awarded 30 nights of bright time on 2.5m Isaac Newton Telescope (INT)
in 1999/06-07 and 2000/09.  The $\sim0.3\Box^{\circ}$ field of view of the Wide
Field Camera (WFC) is ideal for transit searches.  We observed 3 clusters for
the first 20 nights, and one cluster continuously for a further 10 nights.  PSF
photometry was carried out using {\em DAOphot} (Stetson 1987) while
post-processing software was written in-house.  Further details of these
observations and our data reduction procedures can be found in Street et al.
(2002) and Street (2002).  The 20 nights of data on NGC 6819 have been fully
reduced for all 4 CCDs, including over 38,000 stars.
Figure~\ref{fig:RMSfigures} illustrates that we have reached the photometric
precision required to detect hot Jupiter transits.  

\section{Transit Identification} 

We have developed a matched-filter algorithm which can reliably detect transit
events in a reasonable time scale.  Our algorithm initially generates
single-transit model lightcurves for a range of transit durations and epochs.  
These are least-squares fitted to all lightcurves, optimising for transit depth
and out-of-transit magnitude in the process.  The parameters and the 
$\chi^{2}_{1-transit}$ of best-fitting model are recorded and compared with the
$\chi^{2}_{const}$ of the fit of a constant model.   The transit-finder index,
$\Delta\chi^{2}$, is then given by: $\Delta\chi^{2} = \chi^{2}_{const} -
\chi^{2}_{1-transit}$.  

Candidate transit events are identified by plotting $\Delta\chi^{2}$ .vs.
$\chi^{2}_{1-transit}$.  A cut-off is established by fitting a straight line
through the main body of points, then raising it by a specified number of
sigma.  Genuine transits will lie to the upper left of this plot.  All stars
lying above the cut-off are visually examined.   Multi-transit model
lightcurves are then fitted to the selected candidates using a range of
periods, and the parameters and $\chi^{2}$ of the new best-fitting model are
recorded.  The candidate selection procedure is then repeated.  

\section{Results and Future Work}

\begin{figure*}
\centering
\begin{tabular}{cc}
\psfig{file=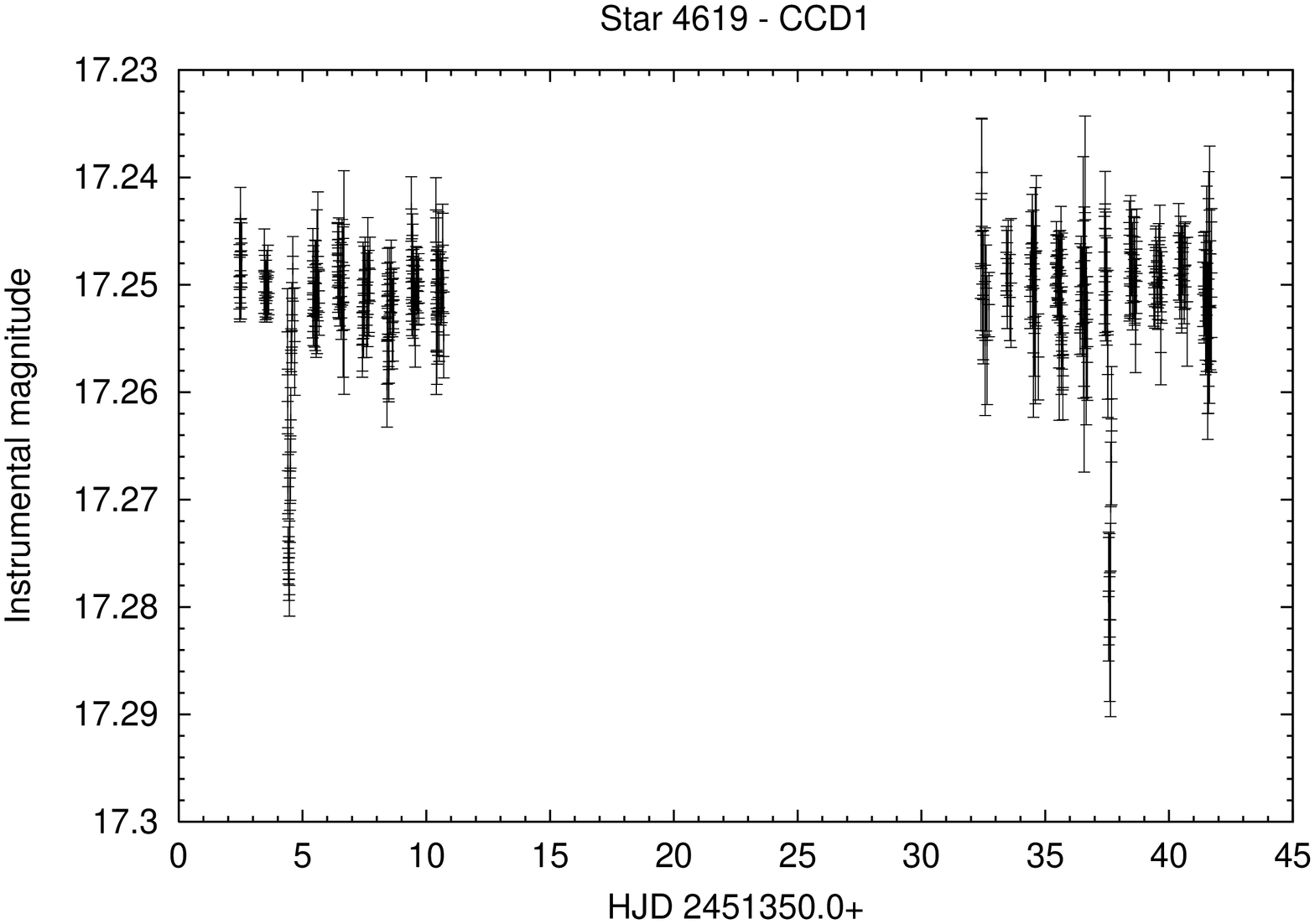,angle=270.0,width=6cm}
&
\psfig{file=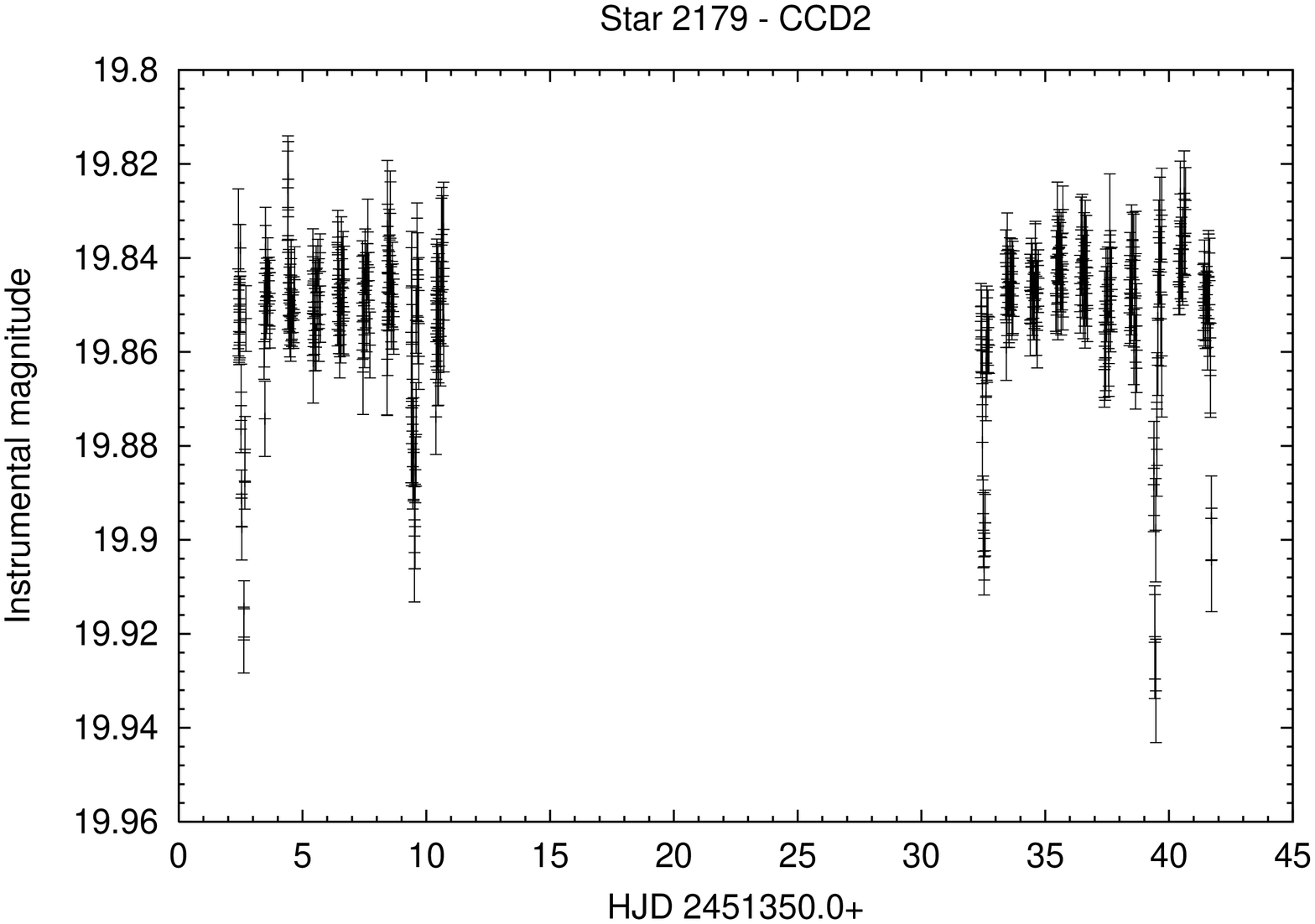,angle=270.0,width=6cm} \\

\psfig{file=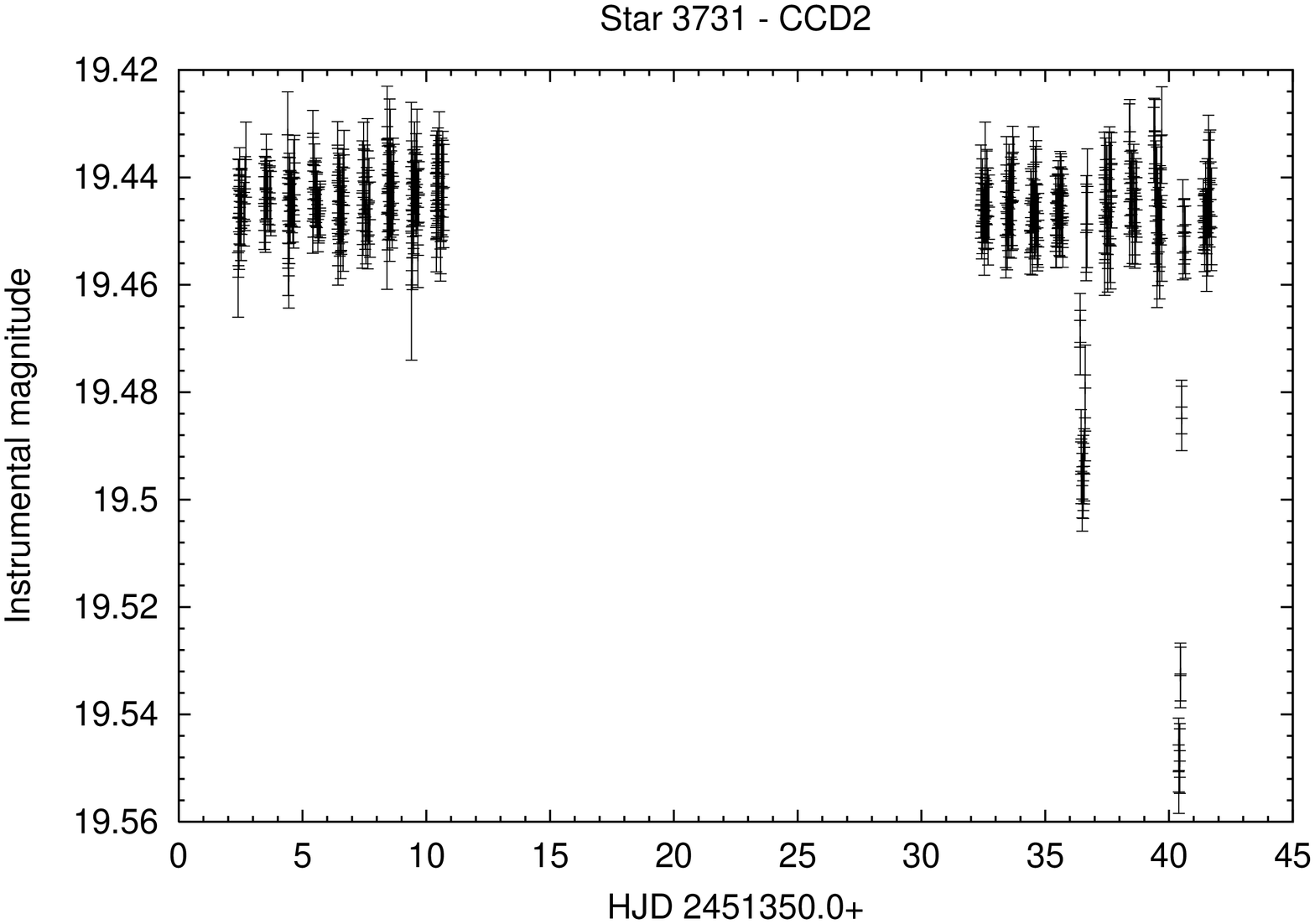,angle=270.0,width=6cm}
&
\psfig{file=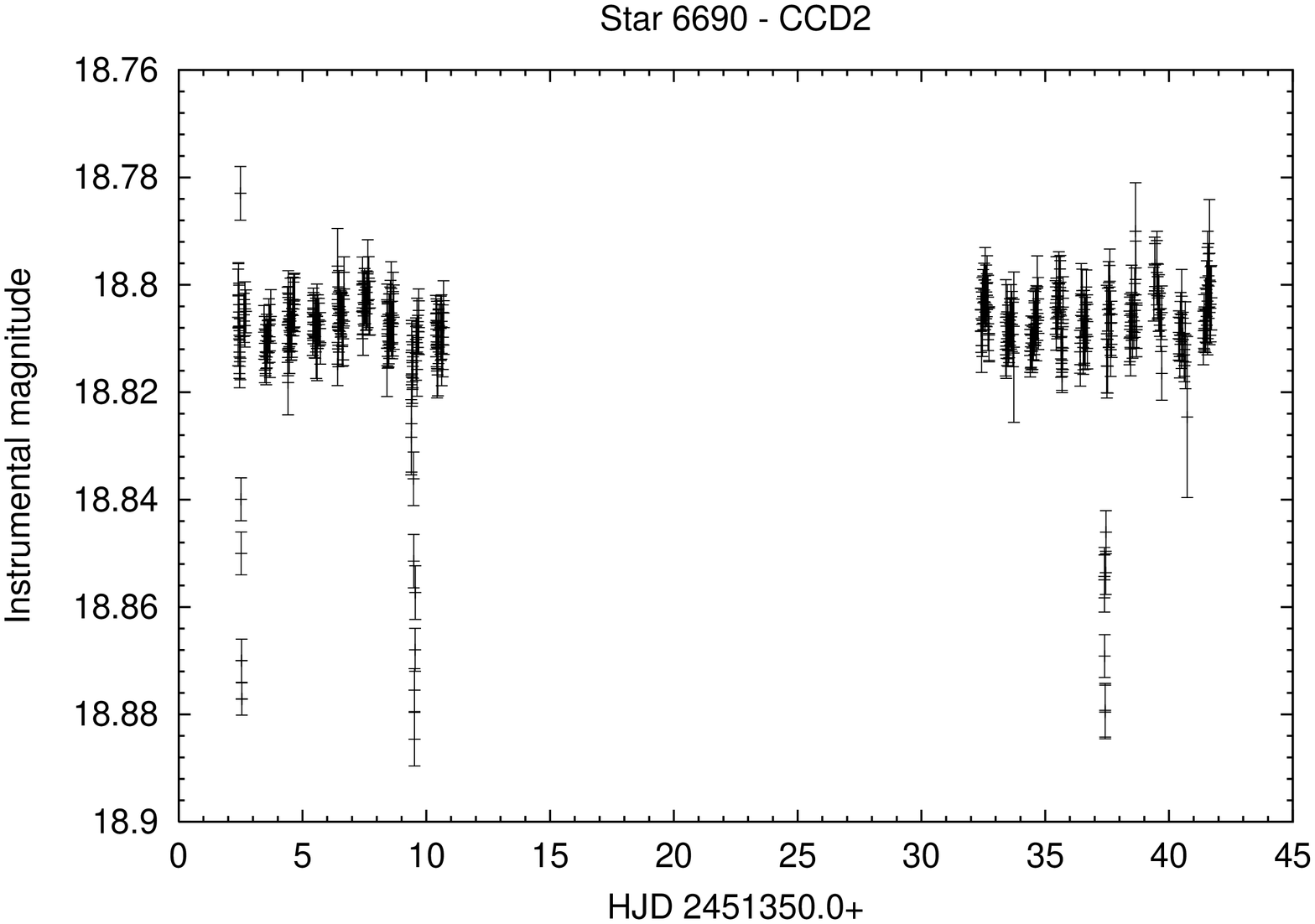,angle=270.0,width=6cm} \\

\psfig{file=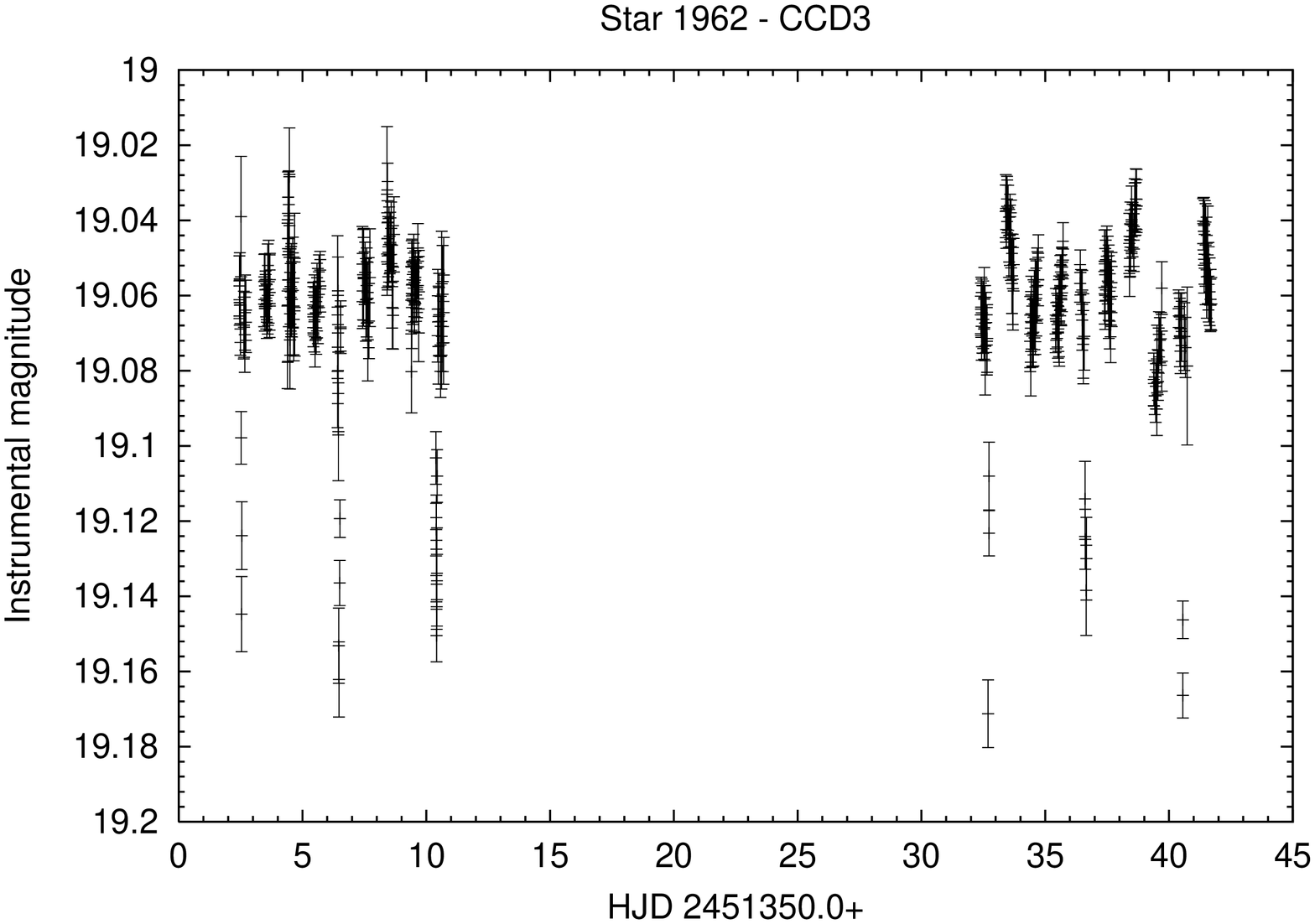,angle=270.0,width=6cm}
&
\psfig{file=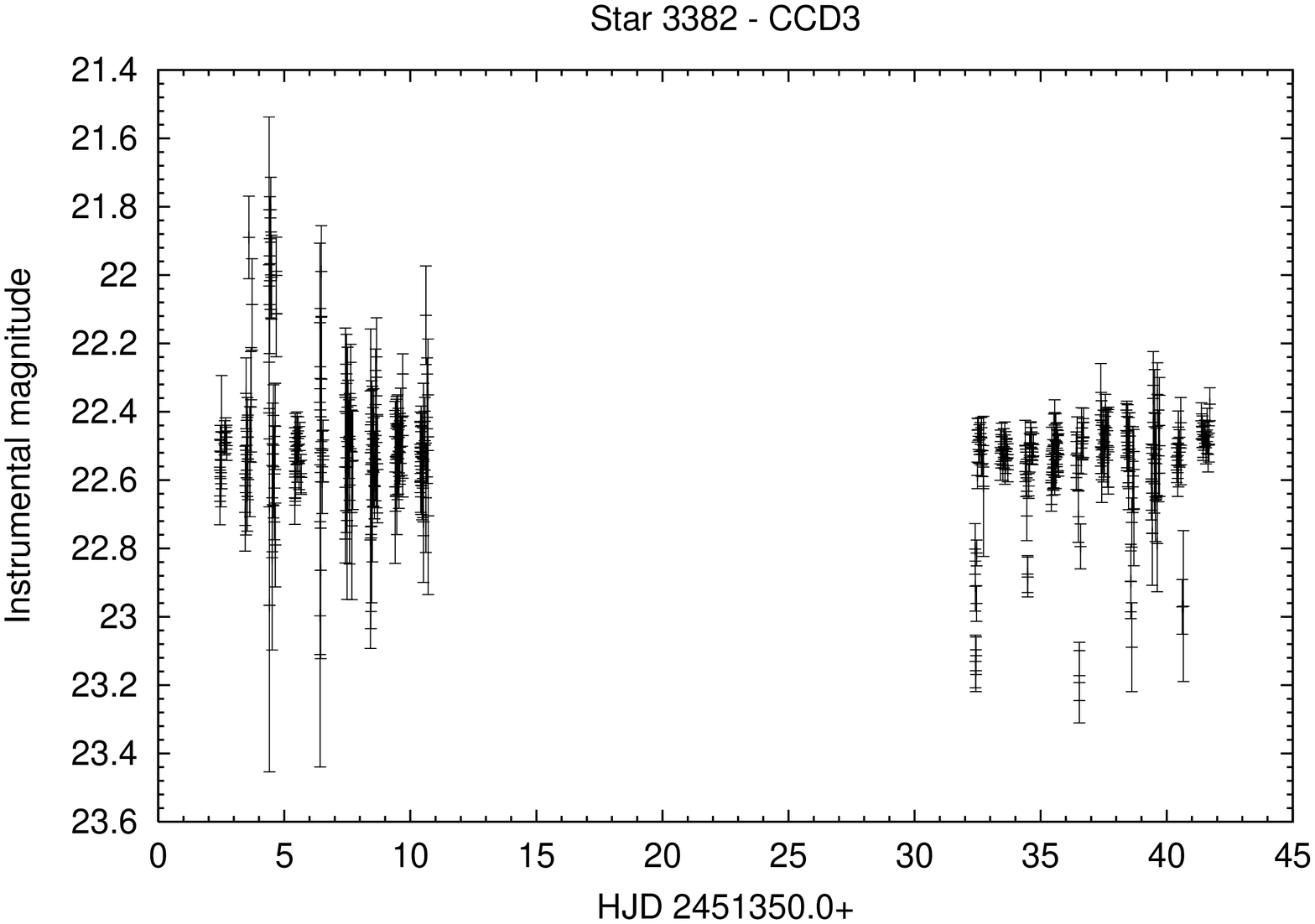,angle=270.0,width=6cm} \\

\psfig{file=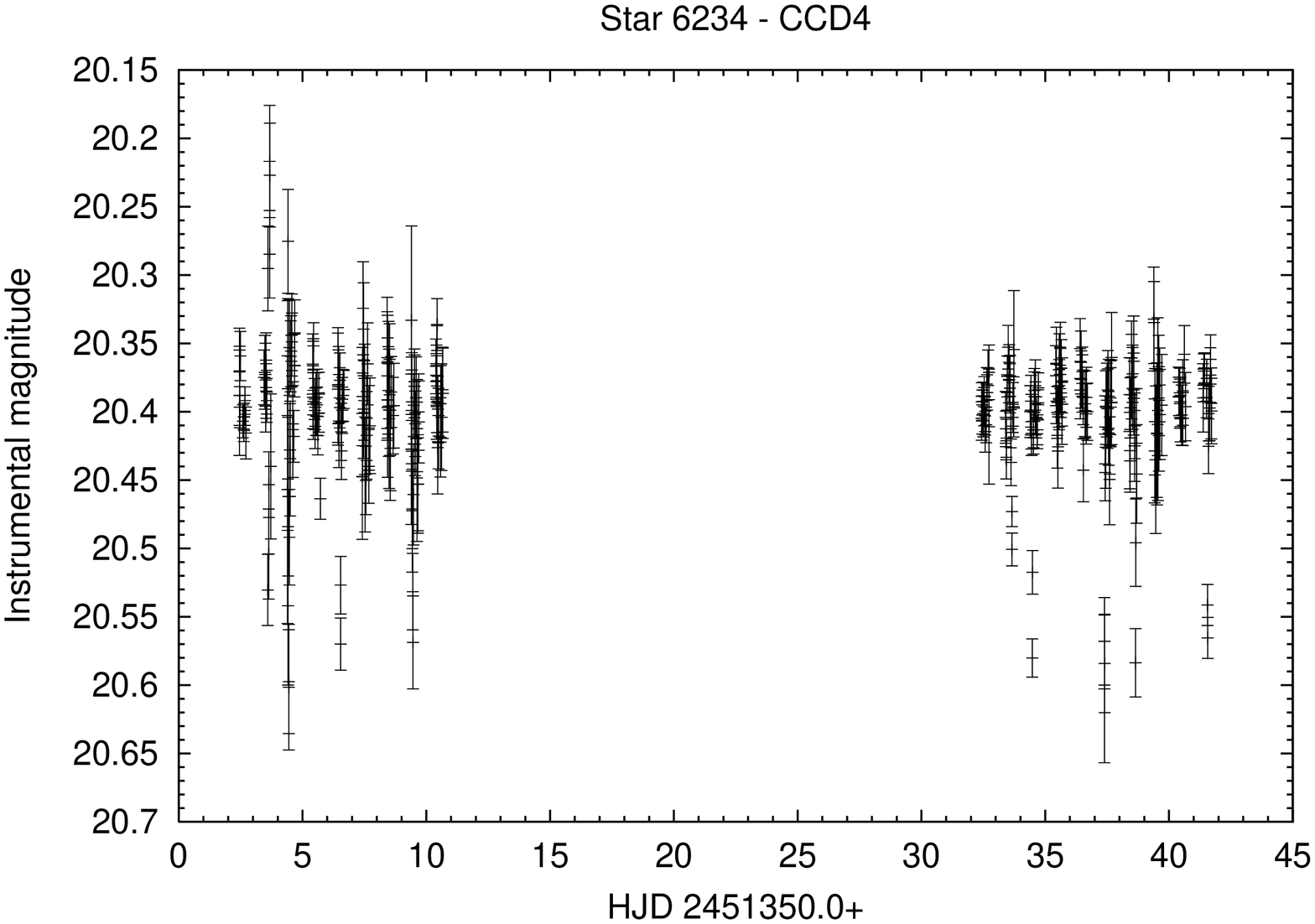,angle=270.0,width=6cm} \\
\end{tabular}
\caption{Lightcurves of stars displaying transit-like events.}
\protect\label{fig:lcs}
\end{figure*}

In this way we have identified 7 stars showing transit-like events (eclipses of
similar amplitude and duration to those of a planetary transit).  The
lightcurves of these stars are shown in Figure~\ref{fig:lcs} and their details
are given in Table~\ref{tab:yield}.

Colour information indicates that in most cases, the companion object is likely
to be a very low-mass star or brown dwarf - interesting objects in their own
right.  However, the eclipses these stars show are very similar to those of
planetary transits, indicating that we can detect such events.  

Our current analysis has some limitations: the algorithm is susceptible to
false alarms caused by blending, stellar activity and scattered data taken in
poor conditions.  We are currently refining our algorithm and investigating
improvements in our reduction to eliminate these effects.  

\begin{table*}
\centering
\caption{The parameters of the stars which show transit-like eclipses.}
\protect\label{tab:yield}
\vspace{5mm}
\begin{tabular}{ccccccc}
\hline
Star	& CCD 	& V mag	 & $V-R$  & Amplitude	& Duration & Period  \\
	&	&	 &	  & (mag)	& (hours)  & (days)  \\
\hline
4619	& 1   	& 16.603 & 0.283  & 0.027     	& 4.0	   & 8.29    \\
2179	& 2   	& 19.130 & 0.396  & 0.05 	& 4.8  	   & 4.6     \\
3731	& 2   	& 18.700 & 0.314  & 0.04-0.09	& 7.2  	   & 8.28    \\
6690	& 2   	& 18.018 & 0.408  & 0.07 	& 2.4	   & 6.98    \\
1962	& 3   	& 15.979 & 0.798  & 0.07 	& 1.8  	   & 1.31    \\
3382	& 3   	& 19.210 & 0.624  & 0.06 	& 2.0  	   & 4.11    \\
6234	& 4   	& 20.340 & 0.695  & 0.04 	& 2.4  	   & 0.42    \\
\hline
\end{tabular}
\end{table*}

\end{document}